\documentclass[pre,aps,twocolumn,superscriptaddress,amsmath,showpacs,floatfix]{revtex4}
\usepackage{graphicx}
\begin{document}
\title{Mode excitation Monte Carlo simulations of mesoscopically large membranes}
\author{Oded Farago}
\affiliation{Department of Biomedical Engineering, Ben Gurion University,
Be'er Sheva 84105, Israel}
\begin{abstract}
Solvent-free coarse grained models represent one of the most promising
approaches for molecular simulations of mesoscopically large
membranes. In these models, the size of the simulated membrane is
limited by the slow relaxation time of longest bending mode. Here, we
present a Monte Carlo algorithm with update moves in which all the
lipids are displaced simultaneously. These collective moves result in
fast excitation and relaxation of the long wavelength thermal
fluctuations. We apply the method to simulations of a bilayer membrane
of linear size $\sim 50\ {\rm nm}$ and show reduction of the
relaxation time by two orders of magnitudes when compared to
conventional Monte Carlo.
\end{abstract}
\maketitle

Biological membranes play a vital role in almost all cellular
phenomena and are fundamental to the organization of the cell. Because
of their remarkable complexity, computer models have become essential
to the understanding of their structure and dynamics. Computer simulations
of lipid and biological membranes can be broadly classified into (i)
atomistic models which are limited in the size and time of problems
they can address by their huge computational workload
\cite{atomistic}, and (ii) coarse grained (CG) models that sacrifice
most of the atomistic details in order to explore larger length- and
time-scales. The field of simplified membrane simulations is more than
20 years old and goes back to the work of Kantor {\em et al.}\/~on
solid tethered membranes \cite{kantor}, which was later extended to
simulations of fluid membrane by considering dynamically triangulated
networks \cite{baumgartner}. A few years later, molecular bead-spring
lipid models were developed to elucidate micelle self-assembly in
aqueous environment \cite{smit}. Recently, a new class of CG molecular
models have been introduced in which bilayer membranes and vesicles
are simulated without direct representation of an embedding solvent
\cite{solventfree}. This is accomplished by constructing
intermolecular force fields that mimic effects of hydration. The
development of implicit-solvent models constitutes an important
advance in large-scale membrane simulations, considering the fact that
the number of solvent particles in explicit-solvent models is
significantly larger than the number of lipids. These models now serve
as platforms for simulations of complexes of membranes with proteins
\cite{brown_proteins,reynwar} and DNA molecules
\cite{farago1}.

Existing implicit solvent CG bilayer models employ an extremely simple
representation of the lipids as short chains consisting of one
hydrophilic bead (representing the head group) and two hydrophobic
beads (representing the hydrocarbon tail), connected to each other by
stiff springs. In earlier works, simulations of membranes consisting
of $N\sim 1000$ lipids have been presented \cite{farago2}. Such
simulations can be easily performed on a commodity PC/workstation. A
membrane patch of 1000 lipids has the linear size of about $L\sim 20\
{\rm nm}$ (taking the area per lipid to be $\sim 0.7\ {\rm nm}^2$),
which is at the small-size end of the mesoscopic regime. Simulations
of larger membranes would require more memory storage and CPU
time. The memory needed for simulations of membranes containing $N\sim
10^4$ lipids is still significantly smaller than the memory available
on a normal PC. The CPU time problem, however, is formidable. For
tensionless membranes, the relaxation time of the longest bending mode
scales as $\tau\sim L^4\sim N^2$ (see Eq.(\ref{eq:qtau})
below). Moreover, for particles interacting via short-range forces
only, the CPU time per Monte Carlo (MC) or Brownian Dynamics time step
scales as $N$. Therefore, the total CPU time of the simulations grows
as $N^3$, from roughly 10 hours for $N=10^3$ (on an AMD Opteron 275
processor running at 2.2 Ghz) to more than a year (!)  for $N=10^4$.
In this work we propose an improved MC scheme that considerably
reduces these enormous computing times and, thus, permit simulations
of membrane-based systems on larger length and time scales.

Recent computer simulations by Reynwar {\em et al.}\/~\cite{reynwar}
demonstrate the slow relaxation problem in membrane simulations: In
this work, the assembly of membrane inclusions by curvature-mediated
interactions was studied \cite{bruinsma_pincus}. Calculating
the interaction between a pair of inclusion requires that the
equilibrium statistics of thermal fluctuations on the scale of the
object pair separation distance are accurately measured. To access the
regime of large separations, Reynwar {\em et al.}\/~employ a CG
implicit-solvent model, which permitted them to simulate a square
membrane of 46080 lipids (the largest membrane patch simulated to
date) with a linear size of about $L\sim 130\ {\rm
nm}$. Unfortunately, the membrane-mediated interactions cannot be
computed at these large spatial scales because the temporal
evolution of the corresponding bending modes is extremely
slow. Therefore, the membrane in this study is decorated with 36
inclusions initially forming a square lattice with spacing $d=L/6\sim
20\ {\rm nm}$, and the calculation of the forces is limited to this
range.

The slowing down of single particle update schemes arises because the
relaxation of large scale fluctuations requires a coordinated movement
of all the lipids over increasingly larger distances. In lattice
membrane models this problem can be solved by using the Fourier
representation of the membrane height field and updating one, randomly
chosen, Fourier amplitude at a time \cite{gouliaev}. (The method was
originally proposed for lattice gauge models \cite{cornell}.) The
efficiency of this method relies the fact that in lattice simulations
each Fourier mode of the membrane represents a single degree of
freedom of the field and, therefore, large variations in their
amplitudes will not be energetically costly and will have reasonable
acceptance probabilities. Such large scale variations are prohibited
in off-lattice molecular models by excluded volume
interactions. Nevertheless, MC algorithms with collective update moves
have been recently proposed for simulations of simple fluids
\cite{liu,maggs}, exhibiting superior performance over conventional MC
schemes. For molecular simulations of membranes and interfaces, we
consider collective moves in which the coordinates of {\em all}\/ the
lipids are simultaneously updated according to the rule:
\begin{equation}
(x,y,z)_{\rm new}=\left(x_{\rm old},y_{\rm old},z_{\rm
old}+\sum_{i=1}^{m}\epsilon_i\cos(\vec{q}_i\cdot \vec{r}+\alpha_i)\right),
\label{eq:scheme}
\end{equation} 
where the sum runs over a set of $m$ modes with wavevectors
$\vec{q}=(2\pi/L)\cdot (n_1,n_2);\ n_1,n_2=0,1,2,\ldots$. This set
includes the modes with the smallest wavenumbers
$n^2=n_1^2+n_2^2=1,2,4,5,8,\ldots$. The random amplitude of the $i$-th
mode in Eq.(\ref{eq:scheme}) is chosen from the interval
$[-\epsilon_i^*,+\epsilon_i^*]$ (the magnitude of $\epsilon_i^*$ is
discussed below), while $\alpha_i$ is a random phase chosen from a
uniform distribution on $[0,2\pi)$. Because this move is reversed by
choosing the set of amplitudes $\{-\epsilon_i\}$, and since the
Jacobian of transformation described by Eq.(\ref{eq:scheme}) is unity,
detailed balance is satisfied when the proposed mapping
Eq.(\ref{eq:scheme}) is combined with Metropolis acceptance rule:
$p({\rm old}\rightarrow {\rm new})={\rm min}(1,\exp(-\beta\Delta E))$,
where $\beta=(k_B T)^{-1}$ is the inverse temperature and $\Delta E$
is the energy difference between the ``new'' and ``old'' states.

The height function of the simulated membrane is calculated by
dividing the area into $M^2=(L/l)^2$ grid cell of size $l$ (comparable to
the width of the membrane), and averaging the height of all the lipids
instantaneously located within each cell \cite{remark}. The Fourier
transform of the discrete height function (defined on the set of
points $\{r_g\}$, each of which is located at the center of a grid cell)
\begin{equation}
h_{\vec{n}}=\sum_{\vec{r}_g}
h\left(\vec{r}_g\right)e^{-2\pi i\,(\vec{n}\cdot\vec{r}_g)/L},
\label{eq:invtransform}
\end{equation}
includes $M^2$ modes, corresponding to $\vec{n}=(n_1,n_2)\ ;\
n_1,n_2=(-M/2)+1,\ldots,M/2$\,. In conventional MC simulations, all
the modes are equally affected by the {\em uncorrelated}\/ move
attempts. Randomly displacing the lipids a vertical distance
$\epsilon$ within a MC time unit, would cause the amplitudes of {\em
all}\/ the Fourier modes (\ref{eq:invtransform}) to change by $(\delta
h_n)^2\sim M^2\epsilon^2=(L/l)^2\epsilon^2$, independently of $n$. At
large scales (small $n$), the behaviour of an undulating membrane can
be described by Helfrich effective surface Hamiltonian which relates
the elastic energy to the local curvature and the bending modulus,
$\kappa$. The power spectrum of the bilayer thermal fluctuations
\cite{safran}
\begin{equation}
\langle |h_{\vec{n}}\,|^2\rangle=\frac{kTL^2}{\kappa l^4|\vec{q}\,|^4}=
\frac{kT L^6}{\kappa \left(2\pi nl \right)^4},
\label{eq:qamplitude}
\end{equation}
strongly depends on $n$. The conventional MC scheme generates
diffusive dynamics in Fourier space, where for each mode the
relaxation time (in MC time units)
\begin{equation}
\tau_n\sim \frac{\langle |h_{\vec{n}}\,|^2\rangle}
{(\delta h_n)^2}=\frac{kT}{\kappa}\frac{L^4}{l^2\epsilon^2
\left(2\pi n\right)^4}.
\label{eq:qtau}
\end{equation}
The relaxation time of the mode with the smallest wavenumber, $n=1$,
grows as a fourth power of the linear size of the system.

The relaxation of long wavelength modes can be accelerated by
introducing collective MC moves which cause larger variations in their
amplitudes. To eliminate the $n^{-4}$ dependence of $\tau_n$ and
ensure that all the modes relax equally fast, we set the interval from
which the random amplitudes in Eq.(\ref{eq:scheme}) are chosen to
satisfy: $\epsilon_i^*=\Delta/n^2$ (see Eq.(\ref{eq:qtau})). The value
of $\Delta$ can be determined empirically, by employing the usual
criterion that the acceptance rate of the moves represented by
Eq.(\ref{eq:scheme}) is approximately half. Notice, however, that
because of the strong decrease of $\epsilon_i^*$ with $n$, significant
improvement in the relaxation times should be expected only for the
longer (also the slower) wavelength modes. Therefore, the sum in
Eq.(\ref{eq:scheme}) can be limited to small wavenumbers while the
relaxation of modes with larger values of $n$ will continued to rely
on single particle moves. The long wavelength modes are efficiently
sampled by the new scheme because the magnitude of $\Delta$ is {\em
independent}\/ of the system size. This can be understood by noting
that the energy cost per unit area of a collective trial move is
$E/L^2\sim CS^2$, where $S\sim(\Delta/L)$ is the induced strain and
$C$ is the relevant elastic modulus. The total deformation energy
$E\sim C\Delta^2$ should be of the order of the thermal energy $k_BT$,
yielding $\Delta^2\sim(k_BT)/C$ which is indeed size-independent. The
collective MC moves cause the amplitudes of the slow modes included in
the sum in Eq.(\ref{eq:scheme}) to change by $|\delta h_n|\sim
M^2\Delta=(L/l)^2(\Delta/n^2)$ (see Eq.(\ref{eq:invtransform})) and,
therefore, their relaxation times scale as
\begin{equation}
\tau_n\sim \frac{\langle |h_{\vec{n}}\,|^2\rangle}
{(\delta h_n)^2}=\frac{kT}{\kappa}\frac{L^2}{\Delta^2
\left(2\pi\right)^4}.
\label{eq:qtau2}
\end{equation}
This time does not increase with decreasing the wavenumber $n$ and,
moreover, grows only as $L^2\sim N$ rather than $L^4\sim
N^2$. Furthermore, a single collective trial move requires the
evaluation of $\mathcal{O}(N)$ (short range) pair-interactions, which
makes them equally CPU time as $\mathcal{O}(N)$ single particle trial
moves. Therefore, the CPU time per MC time unit required in schemes
utilizing $\mathcal{O}(N)$ single-particle and $\mathcal{O}(1)$
collective mode excitation trial moves would scale as $N^2$, which is
superior to conventional MC algorithms whose CPU time grows as $N^3$.

\begin{figure}[t]
\begin{center}
\scalebox{0.35}{\centering \includegraphics{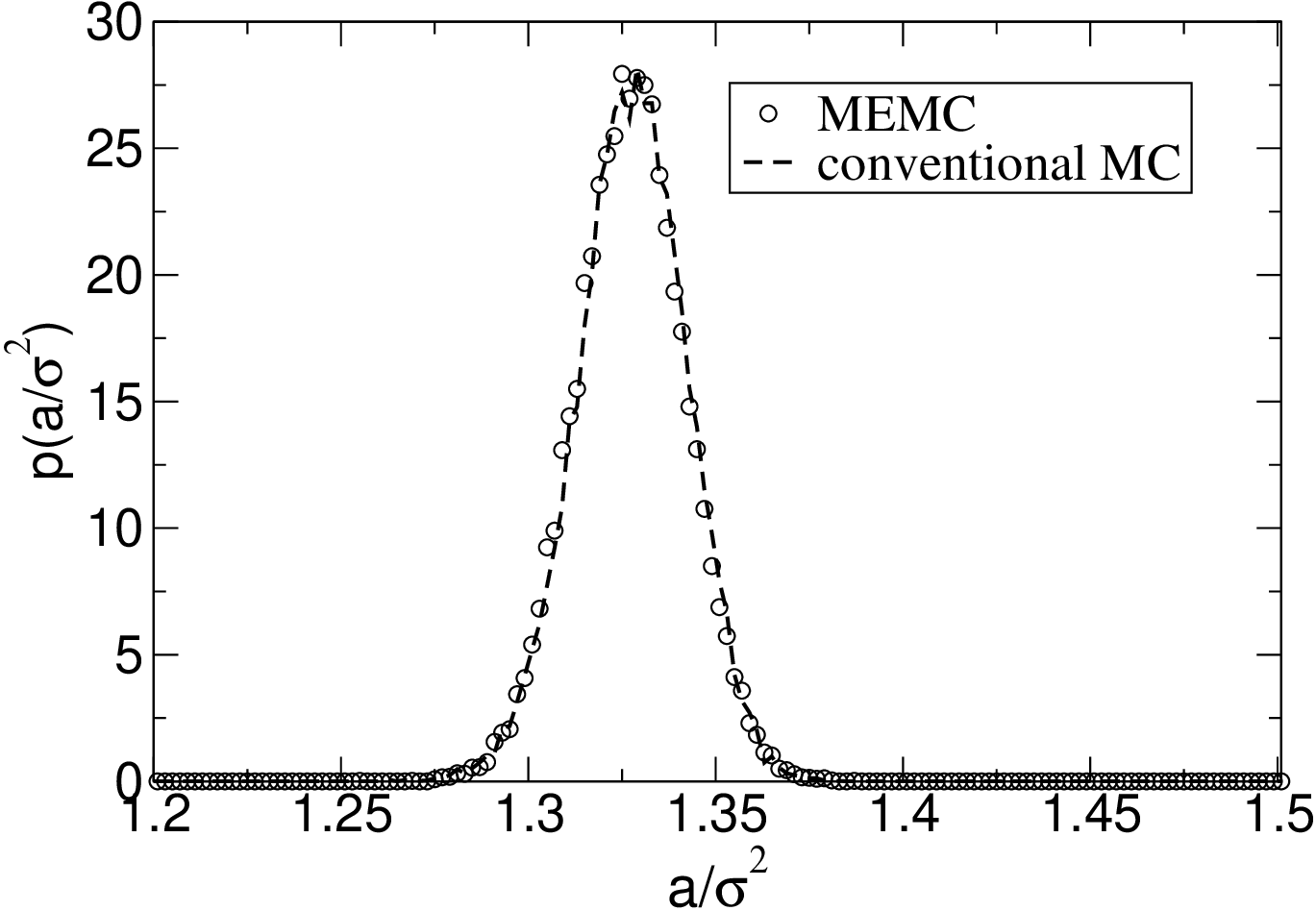}}
\end{center}
\vspace{-0.5cm}
\caption{The normalized distribution functions $p$ of the projected area per 
lipid $a/\sigma^2$, where $\sigma$ is the length parameter of the
bead-bead LJ potential.}
\label{fig:distributions}
\end{figure}

\begin{figure}[t]
\begin{center}
\scalebox{0.35}{\centering \includegraphics{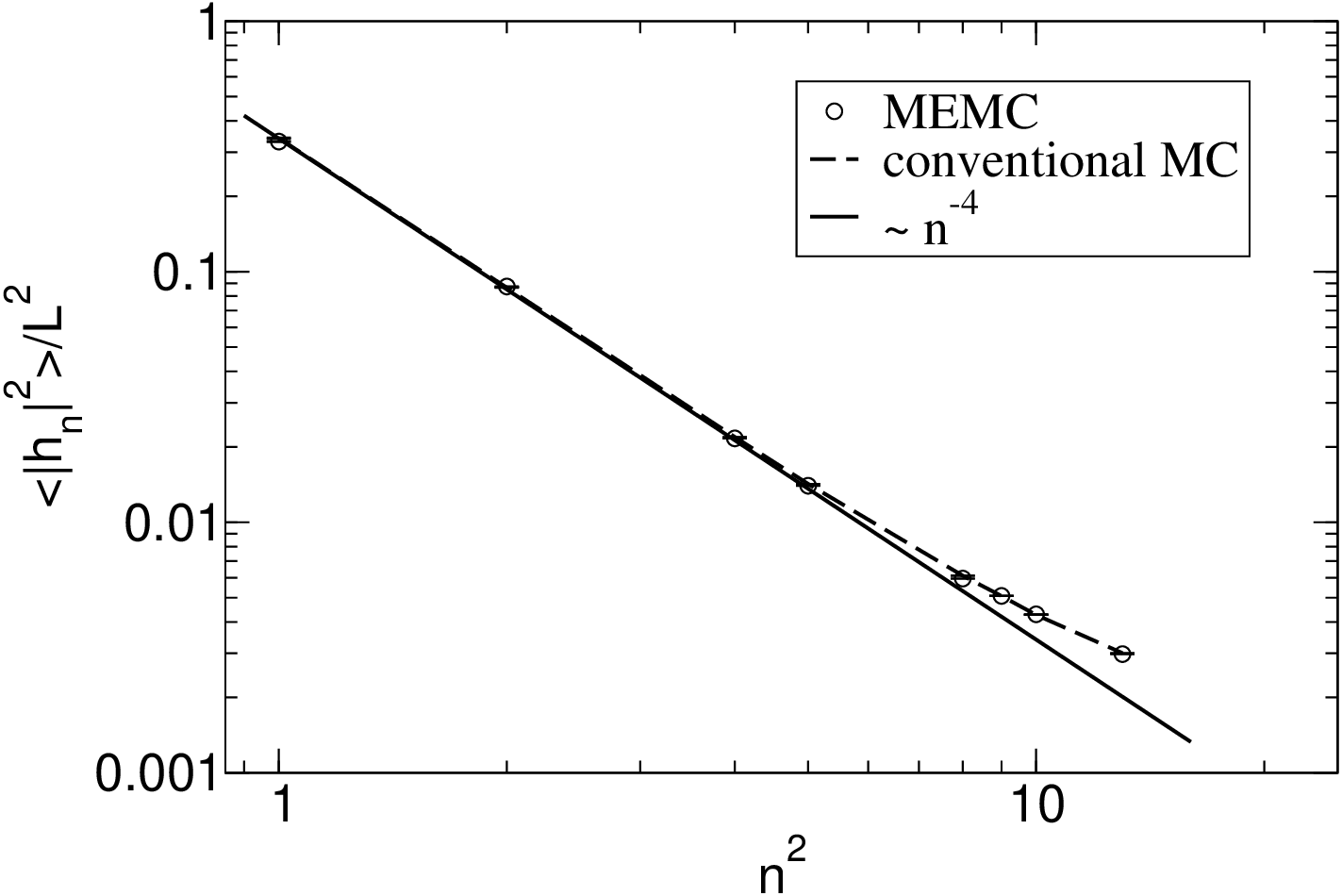}}
\end{center}
\vspace{-0.5cm}
\caption{Fluctuation spectrum of a membrane of $N=1000$ lipids. Results of 
the conventional and improved algorithms are shown by open circles and
dashed line, respectively. The solid line indicates the asymptotic
$\langle |h_n|^2\rangle\sim n^{-4}$ power law.}
\label{fig:spectrum}
\end{figure}

To demonstrate the validity and efficiency of the new algorithm, we
carried out simulations using Reynwar {\em et al.}\/~three-bead lipid
model. The details of the intra- and intermolecular potentials are
given in ref~\cite{cooke}. In our study we set the energy parameter of
the Lennard-Jones (LJ) potential $\epsilon=1.05 k_BT$ and the range of
the attractive tail-tail potential $w_c=1.35\sigma$, where $\sigma$ is
the length parameter of the LJ potential. For this choice of the
parameters, the membrane is in the fluid phase. The intermolecular
interactions were slightly modified from the original model to
eliminate the occasional escape of lipids from the membrane plane,
without affecting the rigidity and fluidity of the membrane. To verify
that the newly proposed {\em mode excitation Monte Carlo}\/ (MEMC)
algorithm works correctly, we used it for MC simulations of square
membranes with $N=1000$ lipids and compared the results to those
obtained by a conventional MC algorithm. The simulations were
conducted in the constant surface tension ensemble
\cite{farago_surface}, at vanishing surface tension. In the
conventional algorithm, each MC time unit consisted of $N$
displacement move attempts of lipids (including changes in the
relative coordinates of the beads), $N$ rotation move attempts, and
two area-changing trial moves. The improved MEMC algorithm included two
additional trial moves per MC time unit in which all the modes with
wavenumbers $n^2\leq 8$ in Eq.(\ref{eq:scheme}) are excited. The
(normalized) distribution functions of the projected area per lipid,
$a=2L^2/N$, obtained from the conventional and improved simulations
are plotted in Fig~\ref{fig:distributions}. Within negligible
computational uncertainties the two distribution functions are
indistinguishable, which confirms that both algorithms generate the
same statistical ensembles. The power spectrum $\langle
|h_n|^2\rangle$ of the height fluctuations is plotted in
Fig.~\ref{fig:spectrum}. The conventional and improved algorithms give
identical results, including the asymptotic $\langle
|h_n|^2\rangle\sim n^{-4}$ power law. From Eq.(\ref{eq:qamplitude})
(setting the mesh size to $l=L/8$), we calculate the bending modulus
of the bilayer $\kappa\simeq 8 k_BT$, in consistency with the values
measured in ref~\cite{cooke}.

\begin{figure}[t]
\begin{center}
\scalebox{0.35}{\centering \includegraphics{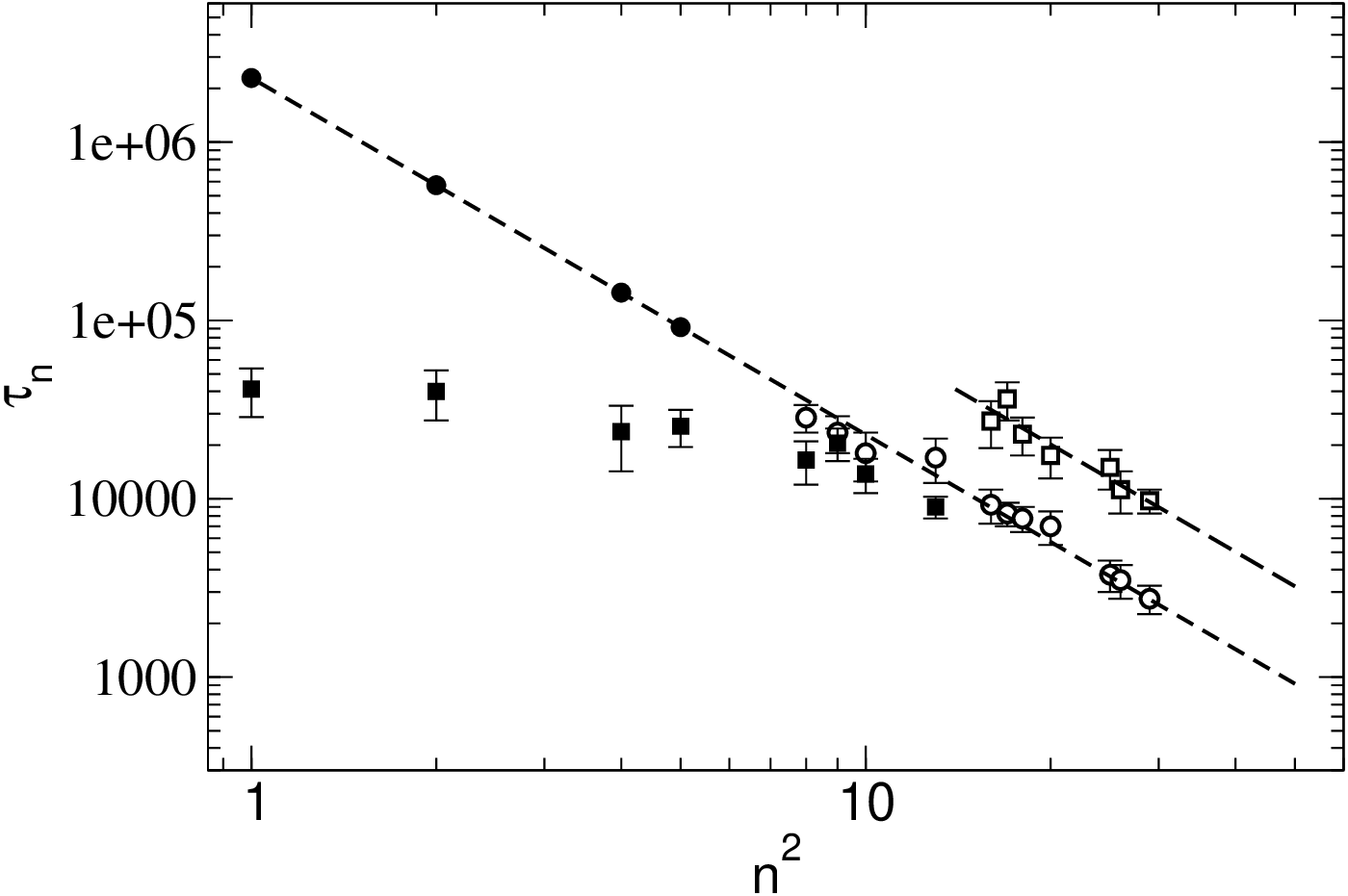}}
\end{center}
\vspace{-0.5cm}
\caption{Relaxation times of undulatory bending modes as a function of 
the wavenumber $n$. Conventional MC results are plotted by circles
(open: results obtained numerically, solid: results evaluated by
extrapolation). MEMC results are plotted by squares (open:
unexcited modes, solid: excited modes). Dashed lines: fits to the
$\tau\sim n^{-4}$ scaling law.}
\label{fig:times}
\end{figure}

Next, we tested the improvement in computational efficiency by
simulating larger membranes consisting of $N=9000$ lipids. The cross
sectional area of the simulation cell was divided into a $16\times 16$
grid and the (discrete) height function was evaluated every 50 MC time
units. The Fourier transform of height function
(\ref{eq:invtransform}) was then computed and the amplitudes off all
the modes with wavenumbers $n^2\leq 29$ were recorded. The relaxation
times were calculated by fitting the time autocorrelation function:
$C_n(\Delta t)=\langle |h_n(t)h_n(t+\Delta t)|\rangle/\langle
|h_n(t)|^2\rangle$ to a double exponential function: $C_n(\Delta
t)=a\exp(-\Delta t/\tau_n^1)+ (1-a)\exp(-\Delta t/\tau_n^2)$. The
double exponential decay has been originally conjectured by Seifert
and Langer \cite{seifert_langer}, and was recently observed in
simulations of Shkulipa {\em et al.}~\cite{shkulipa}. In our
simulations, the dissipation of the bending energy accounts for the
slow relaxation mechanism characterized by $\tau_{n,1}\equiv\tau_n$,
while the smaller relaxation time $\tau_{n,2}$ may be associated with
intermonolayer friction. The latter mechanism was found to play only a
relatively minor role in the decay of all the investigated modes. Due
to the large statistical noise and in order to reduce the cross
correlation between the two relaxation times, the fit to a double
exponential form was limited to time intervals $\tau_{n,2}\ll \Delta
t<\tau_n$. The uncertainties in $\tau_n$ (typically $\pm 20-25\%$)
were determined by comparing the fit results obtained for different
fitting intervals. In the MEMC algorithm, each MC time unit consisted
of (on average): $N$ translations, $N$ rotations, 2 area-changing, and
18 mode (with wavenumbers $n^2\leq 13$) excitation trial moves. The
conventional MC algorithm included only the first three move types
applied with $N:N:2$ proportions; however, each MC time unit of the
conventional algorithm consisted of almost $8N$ trails in order to
make the CPU time per MC time unit of both algorithms the same. The
results of our analysis of the relaxation times are summarized in
Fig.~\ref{fig:times}. The MEMC algorithm eliminates the slowing down
of the long wavelength modes (solid squares), causing them to relax at
very similar rates. The relaxation times of the short modes which are
not excited (open squares) follow the $\tau_n\sim n^{-4}$ power law
(dashed lines), which is also obeyed by the modes when the
conventional MC algorithm is applied (open circles). At small length
scales the MEMC algorithm is almost 4 times slower than the
conventional scheme because each MC time unit of the latter includes
almost 4 times more single particle moves. The relaxation of the total
bilayer area (not shown), which is quite fast, is also slowed down by
a factor of 4. The relaxation times of the long (excited) modes are
considerably reduced and become comparable to the relaxation times of
the longest modes among those which were not excited by the collective
update moves. In comparison to the conventional scheme, the relaxation
of the $n=1$ modes is improved by a factor of about 50, from an
estimated one year of CPU time to less than a week. The simulations
extended over a period of about 10 weeks and, therefore, our estimates
of the long scales slow relaxation times for the conventional MC
algorithm (solid circles) is based on extrapolation of the $\tau_n\sim
n^{-4}$ power law rather than on direct numerical evaluation.

It is interesting to compare the efficiency of the MEMC algorithm with
alternative computational algorithms for constant temperature
simulations. MEMC is clearly more efficient than constant temperature
molecular dynamics (MD) algorithms which at sufficiently large scales
become Brownian in nature and effectively behave like conventional MC
simulations \cite{grest_kremer}. Improved relaxation behavior is
achieved when the MD simulations are run with a momentum-conserving
thermostat \cite{dpd} that, on long length and time scales, reproduce
the correct hydrodynamic behavior $\tau\sim L^3$ \cite{granek}. When
the CPU time per time step is considered, one finds that the
computational complexity of such simulations grows as $N^{2.5}$. This
is better than conventional MC and MD but still inferior to the MEMC
algorithm whose complexity grows as $N^2$. In the lattice membrane
simulations \cite{gouliaev}, the CPU time per MC time step grows as
$N^2$ (since there are $\mathcal{O}(N)$ Fourier modes and the
variation of each is a collective move that requires the calculation
of $\mathcal{O}(N)$ interaction terms), which makes it comparable to
MEMC simulations of tensionless membranes when each lattice point
represents a microscopic area element of the membrane. However, when
the membrane is under tension or in the presence of an external
harmonic potential, the power spectrum for small wavenumbers are given
by $\langle |h_{\vec{n}}\,|^2\rangle\sim n^{-2}$ and $\langle
|h_{\vec{n}}\,|^2\rangle\sim n^{0}$, respectively
\cite{safran}. Repeating the argument that leads from
Eq.(\ref{eq:invtransform}) to Eq.(\ref{eq:qtau2}), one finds
$\tau_n\sim L^0$ rather than $\tau_n\sim L^2$ in Eq.(\ref{eq:qtau2})
and, therefore, the required CPU time for MEMC simulations of such
membranes would grow only linearly with $N$. This demonstrates that
the MEMC algorithm is asymptotically faster not only than other
algorithms for continuum (molecular) membrane simulations, but also
than the Fourier MC algorithm for lattice simulations.

To summarize, we introduce an improved MC algorithm for simulations of
mesoscopically large membranes. The new algorithm utilizes collective
update moves that lead to fast excitation and relaxation of the long
wavelength bending modes. The slow relaxation of these modes in
conventional MC and MD schemes is the most severe constraint that
limits the size of the simulated membranes in solvent-free coarse
grained models. The efficiency of the new algorithm is demonstrated by
simulations of a membrane patch of 9000 lipids, where a 50-fold
decrease in the relaxation time was measured as compared to a
conventional MC algorithm with only single particle moves. Implicit
solvent bilayer models combined with improved sampling techniques,
such as the mode excitation algorithm presented here, can serve as the
basis for large scale CG simulations of complexes of bilayer membranes
with additional biological components.


\begin{thebibliography}{99}
\bibitem{remark} Averages were calculated separately for each monolayer of 
the membrane, and the average of these two averages was taken.
\end{thebibliography}
\end{document}